\documentclass{aa}
\usepackage{graphics}
\def\a{{$\alpha$}}

\def\gsnr{{G~34.7-0.4}}
\newcommand{\h}{$^{\rm h}$}
\newcommand{\m}{$^{\rm m}$}
\newcommand{\s}{$^{\rm s}$}
\newcommand{\dd}{$\delta$}
\newcommand{\ha}{\rm H$\alpha$}
\newcommand{\hbeta}{\rm H$\beta$}

\newcommand{\HI}{\ion{H}{i}}
\newcommand{\hnii}{{\rm H}$\alpha+[$\ion{N}{ii}$]$}
\newcommand{\nii}{$[$\ion{N}{ii}$]$}
\newcommand{\sii}{$[$\ion{S}{ii}$]$}
\newcommand{\oi}{$[$\ion{O}{i}$]$}

\newcommand{\oii}{$[$\ion{O}{ii}$]$}
\newcommand{\oiii}{$[$\ion{O}{iii}$]$}

\newcommand{\et}{et al.}
\newcommand{\flux}{$10^{-17}$ erg s$^{-1}$ cm$^{-2}$ arcsec$^{-2}$}
\newcommand{\dens}{\rm cm$^{-3}$}
\newcommand{\sdens}{\rm cm$^{-2}$}
\newcommand{\vel}{\rm km s$^{-1}$}
\begin{document}

%   \thesaurus{08     % A&A Section 8: Diffuse matter in space
%              (09.07.01;  % ISM : general,
%		09.19.2;
%		09.09.1)} % Superona remnants
%
\title{The faint supernova remnant \gsnr\ (W44)}
\author{F. Mavromatakis\inst{1}
\and P. Boumis\inst{2}
\and C. D. Goudis\inst{2,3}}
\offprints{F. Mavromatakis,\email{fotis@physics.uoc.gr}}
\authorrunning{F. Mavromatakis}
\titlerunning{Optical observations of \gsnr}
\institute{
University of Crete, Physics Department, P.O. Box 2208, 710 03 Heraklion, 
Crete, Greece 
\and
Institute of Astronomy \& Astrophysics, National Observatory of Athens, 
I. Metaxa \& V. Pavlou, P. Penteli, 15236 Athens, Greece
\and
Astronomical Laboratory, Department of Physics, University of Patras, 26500
Rio--Patras, Greece
}
\date{Received 11 March 2003/Accepted 23 April 2003}

\abstract{
Flux calibrated images of the known supernova remnant \gsnr\ in basic optical 
emission lines are presented. The low ionization images show a relatively
flat flux distribution. The diffuse and patchy morphology of the detected
optical emission may indicate the presence of turbulent magnetic fields. 
Typical observed \hnii\ fluxes are $\sim$8$\cdot$\flux, while the \sii\ fluxes
are lower around 4$\cdot$\flux. Emission in the medium ionization line of \oiii
5007 \AA\ is not detected within our sensitivity limits, probably due to the heavy
extinction towards the remnant. The long--slit spectra reveal strong \sii\ and
\nii\ emission relative to \ha\ and moderate \oi 6300 \AA\ emission. Shock 
velocities in the range of 110--150 \vel\ and low electron densities are
estimated. Archival MSX infrared data show emission in the south and west 
areas of the remnant matching rather well the optical and radio emission.       
\keywords{ISM: general -- ISM: supernova remnants
-- ISM: individual objects: G 34.7-0.4}
}
\maketitle
\section{Introduction}
The radio emission from \object{G34.7-0.4} (\object{W44}) is non--thermal 
($\alpha\sim\ $--0.4; I $\sim\nu^{\alpha}$) and displays a shell like morphology.
The shell displays an ellipsoidal shape with the major axis oriented in 
the south--east to north--west direction and an extent of
$\sim$35\arcmin$\times$26\arcmin\ 
(e.g. Jones \et\ \cite{jon93}, Giacani \et\ \cite{gia97}).
The strongest filamentary radio emission is observed in the eastern half of 
the remnant, while  
Kundu \& Velusamy (\cite{kun72}) found that the projected magnetic field 
distribution is quite uniform in the northeast areas of the remnant. 
The X--ray images of the object reveal a centrally brightened morphology 
with the X--ray emission bounded by the radio emission (Rho \et\ \cite{rho94}). 
Temperature measurements of the hot gas lie in the range of 
0.7--1.0 keV, while the neutral hydrogen column 
density is $\sim$10$^{22}$ \sdens\ (Harrus \et\ \cite{har97}, Jones \et\
\cite{jon93}). The 21 cm line observations of Koo \& Heiles (\cite{koo91}) 
revealed the presence of an \HI\ shell associated with W44 expanding with 
a velocity of $\sim$150 \vel, and carrying $\sim$8$\cdot$10$^{44}$ erg of kinetic 
energy. 
\par
Narrow band optical images of the remnant have been presented by Rho \et\
(\cite{rho94}) and Giacani \et\ (\cite{gia97}) in the emission lines of \hnii\ 
and \sii. These images show the diffuse nature of the optical emission, its
low surface brightness and a satisfactory degree of correlation with the radio
emission. Interest on this remnant was further raised by the discovery of a 267
ms pulsar at a distance of $\sim$3 kpc and a characteristic spin--down age of
20000 yr (Wolszczan \et\ \cite{wol91}). Subsequent deep X--ray observations by
ASCA and Chandra in the vicinity of the pulsar lead to the discovery of 
an extended nebulosity. Its spectrum is hard and can be described by a 
steep power--law model (Harrus \et\ \cite{har96}, Petre \et\ \cite{pet02}).
The emission around the pulsar is considered to originate from a synchrotron 
nebula powered by \object{PSR B1853+01}. 
\par
Our current view of W44 consists of a remnant in the radiative phase of its
evolution with a radius of $\sim$12 pc at a distance of $\sim$2.5 kpc (Cox \et\
\cite{cox99}). Thermal conduction has, most likely, altered the interior 
density and temperature distributions from those expected by an 
adiabatic remnant. The progenitor star exploded $\sim$20000 yr ago releasing
$\sim$10$^{51}$  erg into an interstellar medium (ISM) of relatively high
density ($\sim$3--6 \dens). The ISM density is not uniform but displays a 
significant
gradient perpendicular to the major axis of the remnant with a scale height of
$\sim$22 pc (Cox \et\ \cite{cox99}, Shelton \et\ \cite{she99}). 
\par
In this work we present flux calibrated images in the optical emission 
lines of \hnii,
\sii\ and \oiii\ as well as long--slit spectra at three different locations of
W44. In Sect. 2 we provide information about the observations and data 
analysis, while in Sect. 3 and 4 the results of the imaging and spectroscopic 
observations are presented. Finally, we discuss the properties of the supernova
remnant and the environment into which lies in Sect. 5. 
  \begin {figure}
   \resizebox{\hsize}{!}{\includegraphics{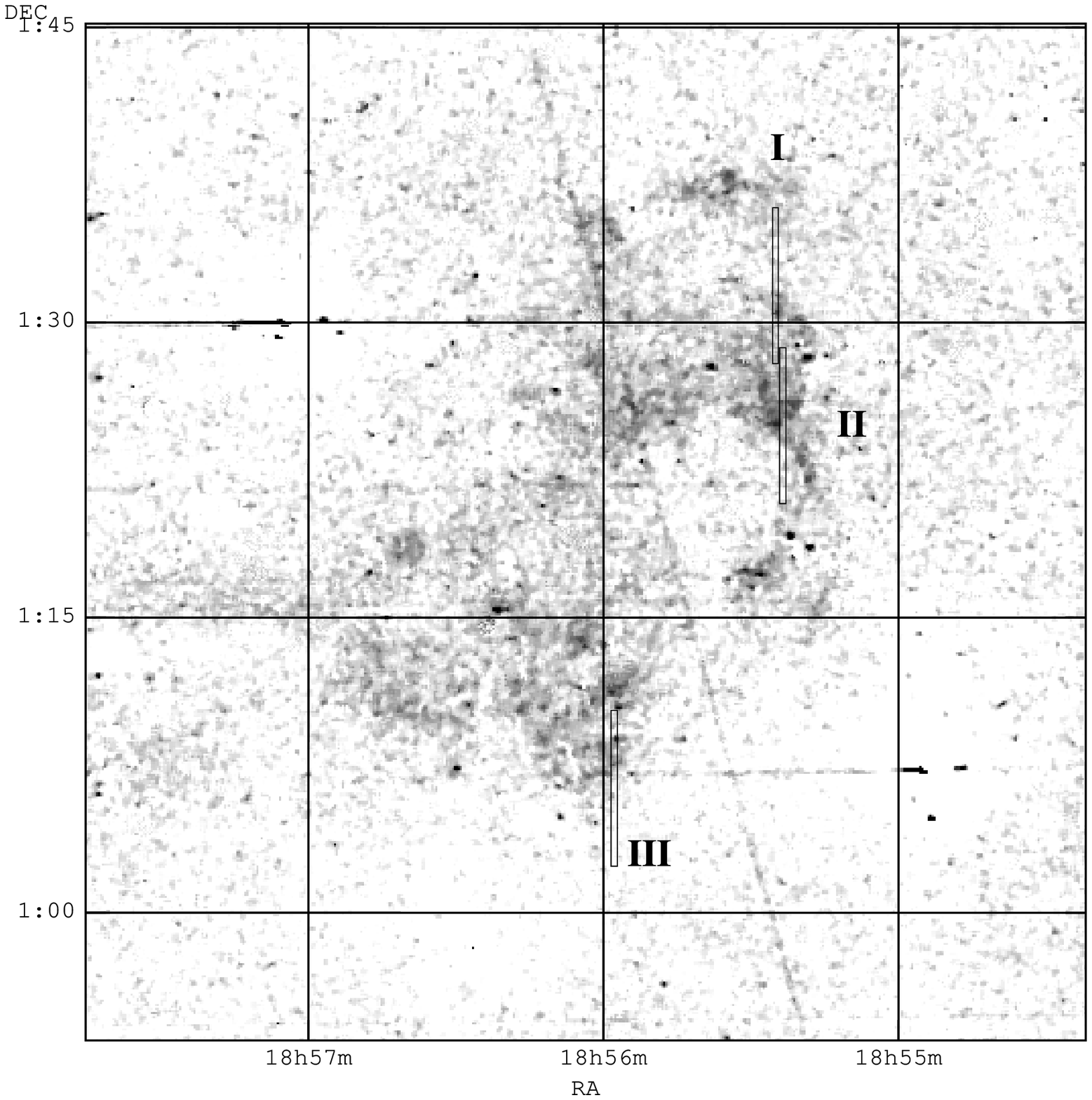}}
    \caption{
     The diffuse and patchy nature of the emission from \gsnr\ 
     is the major characteristic of the \hnii\ image. 
     The shadings run linearly from 0 to 20 $\cdot$ \flux\ and the 
     long rectangles mark the approximate projection of the slits 
     on the sky. The image has been smoothed to suppress the 
     residuals from the imperfect continuum subtraction.
     The line crossing the field in, roughly, the south--north 
     direction is due to a satellite pass. 
     The line segments seen near over-exposed stars in this 
     figure and the next figures are due to the blooming effect.
      } 
     \label{fig01}
  \end{figure}
%--------------------------------------------------------
  \begin {figure}
   \resizebox{\hsize}{!}{\includegraphics{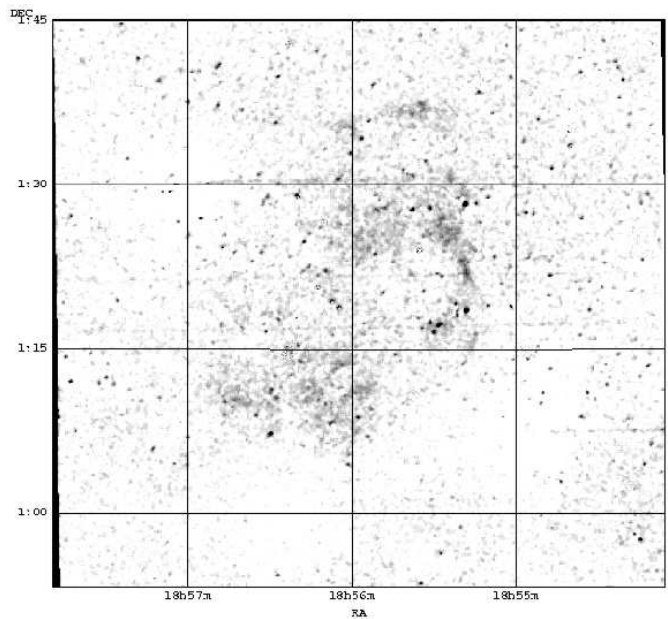}}
    \caption{ The field of \gsnr\ imaged with a filter that isolates 
     the sulfur lines at 6716 and 6731 \AA. 
     The image has been smoothed to suppress the residuals 
     from the imperfect continuum subtraction, while the
     shadings run linearly from 0 to 10 $\cdot$ \flux. 
     } 
    \label{fig02}
  \end{figure}
\section{Observations}
%
%----------------
  \begin{table}
      \caption[]{Spectral log}
         \label{spectra}
\begin{flushleft}
\begin{tabular}{lllll}
            \noalign{\smallskip}
\hline
	Slit centers &  & Exp. times$^{\rm a}$ & \cr
\hline
	 $\alpha$ & $\delta$ & (No of spectra$^{\rm b}$)  &  \cr
 \hline 
18\h55\m25\s & 01\degr32\arcmin04\arcsec 	& 7200 (2) \cr
 \hline
18\h55\m59\s & 01\degr05\arcmin18\arcsec 	& 7800 (2)\cr
 \hline
18\h55\m24\s & 01\degr24\arcmin44\arcsec 	& 7800 (2) \cr
 \hline
\end{tabular}
\end{flushleft}
${\rm ^a}$ Total exposure times in sec\\\
${\rm ^b}$ Number of spectra obtained \\\
   \end{table}
 \begin{table*}
        \caption[]{Relative line fluxes}
         \label{sfluxes}
         \begin{flushleft}
         \begin{tabular}{lllllllll}
     \hline
 \noalign{\smallskip}
                &Area I  &                  & Area II   &       &Area  III  \cr
\hline
Line (\AA) & F$^{\rm a,b}$ & I$_{\rm corr}^{\rm c}$ & F$^{\rm a,b}$ & 
I$_{\rm corr}^{\rm c}$ &  F$^{\rm a,b}$  & I$_{\rm corr}^{\rm c}$  \cr
\hline
4861 \hbeta\   & --      & 100    & --       & 100     & 12 (3) &  100 \cr
\hline
4959 \oiii$_1$ & --      & --    & --        & --  & --     &  -- \cr
\hline
5007 \oiii$_2$ & 10$^{\rm d}$ (3)  & 901   & --        & --  & --      & -- \cr
\hline  
6300 \oi$_1$  & 23 (6)   & 108   & 26 (5)   & 123  & 59 (7) & 279 \cr
\hline
6364 \oi$_2$  & 7 (4)    & 25    & 8 (2)    & 36  & 28 (4)  & 119 \cr
\hline
6548 \nii$_1$ & 28 (5)  & 85    & 26 (8)   & 80   &20 (4)  & 61 \cr
\hline
6563 \ha\     & 100 (18) & 300   & 100 (30)  & 300  &100 (19)& 300 \cr
\hline
6584 \nii$_2$ & 86 (15)  & 249   & 84 (25)   & 242  &81 (15) & 234 \cr
\hline
6716 \sii$_1$ & 80 (15)  & 186   & 97 (29)   & 228  &75 (16) & 176 \cr
\hline
6731 \sii$_2$ & 59 (12)  & 134   & 71 (22)   & 162  &50 (10) & 114 \cr
\hline
\hline
Absolute \ha\ flux$^{\rm e}$ & 3.8       &       & 4.1  &  & 2.9  &   &  \cr
\hline
\sii\ /\ha\           & 1.39 [0.11]&1.07 & 1.68[0.07]& 1.29& 1.25[0.10]& 0.97 \cr
\hline
\nii/\ha\            & 1.15 [0.09]&1.11 & 1.12[0.06]& 1.08& 1.08[0.09]&1.04 \cr
\hline
\sii$_1$/\sii$_2$    & 1.36 [0.15]&1.38  & 1.36[0.08]& 1.40& 1.51[0.18]&1.55 \cr
\hline
n$_{\rm e}$ (1$\sigma$) & $<$ 220 \dens&     & $<$ 140 \dens &      & $<$ 200 \dens &      \cr 
\hline
\hline 
\end{tabular}
\end{flushleft}
 ${\rm ^a}$ Fluxes uncorrected for interstellar extinction and relative to
 F(\ha)=100

${\rm ^b}$ Listed fluxes are a signal to noise weighted
average of the individual fluxes

$^{\rm c}$ Corrected for interstellar extinction assuming E(B--V)=3.28 or c=4.94
 
$^{\rm d}$ Detected only in one of the two spectra

$^{\rm e}$ In units of \flux\ 

${\rm }$ Numbers in parentheses represent the signal to noise ratio 
of the quoted fluxes \\
${\rm }$ 1$\sigma$ errors are given in square brackets 
\end{table*}
%-------------------------------------------------------

\subsection{Optical images}
The supernova remnant W44 was observed with the 0.3 m Schmidt--Cassegrain 
telescope at Skinakas Observatory, Crete, Greece on July 7 and 
August 7--8, 2002. 
The observations were performed with a 1024 $\times$ 1024 Thomson CCD 
which provides a 70\arcmin\ $\times$ 70\arcmin\ field of view and an 
image scale of 4\arcsec\ per pixel. The remnant was observed through 
the \hnii, \sii, and \oiii\ filters. 
Astrometric solutions were calculated for all data frames using the HST 
Guide star catalogue (Lasker \et\ \cite{las99}) and were subsequently projected 
to a common origin on the sky before any arithmetic operation. 
The equatorial coordinates quoted in this work refer to epoch 2000.  
\par
Standard IRAF and MIDAS routines were employed for the reduction of the data. 
All frames were bias subtracted and flat-field corrected using a series of
well exposed twilight flat--fields. 
The absolute flux calibration was performed through observations of a series 
of spectrophotometric standard stars (HR5501, HR7596, HR7950, HR9087 and 
HR8634; Hamuy \et\ \cite{ham92}, \cite{ham94}). 
The digital data were retrieved from the European Southern Observatory
(ESO) and were folded with the appropriate filter transmission curve,
which is known for the specific focal ratio of the 
0.3 m telescope.For the given airmass, exposure time of each calibration 
star and the instrumental setup, it is possible to predict the expected 
count rate. 
A least squares fit for the expected and observed count rates allows us to
calculate  the zero point magnitude and the atmospheric extinction in the 
specific filter. These coefficients are then applied to the data frames 
for absolute flux calibration (see also Jacoby \et\ \cite{jac87}).
\subsection{Optical spectra}
Long--slit spectra were obtained on July 6, 7, and 8, 2002 with the 1.3 m 
Ritchey--Cretien telescope at Skinakas Observatory. 
The data were taken with a 1300 line mm$^{-1}$ grating 
and a 800 $\times$ 2000 SITe CCD covering the range of 4750 \AA\ -- 6815 \AA.
The projected slit width on the sky is 7\farcs7, while its length is 7\arcmin.9.
The coordinates of the slit centers,
the number of available spectra from each location and the total exposure 
times are given in Table~\ref{spectra}. 
The spectrophotometric standard stars HR5501, HR7596, HR7950, HR9087, 
and HR8634 were observed for the absolute calibration of the source spectra.  
The spectra were reduced with the {\it IRAF} software and specifically, 
with the packages {\it twodspec}, {\it longslit} and {\it apextract}. The
sufficient coverage in airmass of the calibration stars allowed to determine 
both the sensitivity function of the instrument response and the dependence 
of the airmass on the wavelength. Subsequently, these corrections were applied 
to the data spectra.
\section{The imaging observations}
\subsection{The \hnii\ and \sii\ line images}
The low ionization images covering the \ha, \nii, and \sii\ lines 
(Fig. \ref{fig01} and \ref{fig02}) display a diffuse and patchy morphology 
already noted by Giacani \et\ (\cite{gia97}) and Rho \et\ (\cite{rho94}). 
The flux 
in each of the  \hnii\ and \sii\ images appears to have a smooth distribution over
the extent of the remnant. Since our images are flux calibrated, we can measure
an average \hnii\ flux of $\sim$8$\cdot$\flux\ over several locations of the
remnant with faint diffuse emission emitting at a level of 3$\cdot$\flux.  
The estimated \ha\ flux is $\sim$4$\cdot$\flux\ since the \nii 6548, 6584 \AA\
lines contribute equally well in the \hnii\ filter and we have assumed that 
the \ha\ and \nii\ fluxes are of comparable strength (e.g. Fesen \et\
\cite{fes80}, Mavromatakis \et\ \cite{mav02}; see also \S 4).
Measuring the observed sulfur flux over the same locations we find an average 
value of 4$\cdot$\flux, while the faintest diffuse emission amounts to a flux 
of $\sim$2$\cdot$\flux. However, the two sulfur lines do not contribute equally
in the filter bandpass and in order to account for this effect we assume a 
\sii 6716\AA/\sii 6731 \AA\  ratio of 1.4, appropriate for middle aged remnants.
Consequently, the average total sulfur flux is estimated around
6$\cdot$\flux. Although the observed emission in the \hnii\ and \sii\ lines is
weak in terms of absolute flux, the sulfur flux is quite strong relative to
\ha. Using the calibrated images to estimate the observed \sii/\ha\ ratio, it is
found that it ranges from 0.9--1.8. 
These values are in agreement with those measured in the available long--slit 
spectra (\S 4).  
\subsection{The \oiii\ image}
The final \oiii\ image does not show any emission from the remnant and thus, it
is not shown here. The total exposure time of 7200 s and the hardware 
configuration used did not allow the detection of possible \oiii 5007 \AA\ 
emission from W44 and the 3$\sigma$ upper limit set is 1.5$\cdot$\flux\ over 
its whole area. This upper limit is consistent with the spectroscopic 
measurements.
\section{The long--slit spectra from \gsnr}
Long--slit spectra were obtained for the first time from this faint remnant.
Approximate slit positions are shown in Fig. \ref{fig01} where it is seen that
locations I, II, and III mainly focus on the brighter areas of the remnant 
in the south, and north--west. Especially, location II was chosen 
due to the stronger \sii\ emission in that area. The measured fluxes from 
individual lines along with some derived quantities are given in 
Table \ref{sfluxes}. 
Since a reliable Balmer decrement measurement towards \gsnr\ is not
available, we adopt the statistical relation 
\begin{equation}
{\rm N_{H}} = 5.4~(\pm 0.1) \cdot\ 10^{21}~{\rm E(B-V)}~{\rm cm}^{-2},
\end{equation}
given by Predehl and Schmitt (\cite{pre95}) to determine the color 
excess. Assuming a hydrogen column density of 1.77 $10^{22}$ \sdens\ 
(Tsunemi \& Enoguchi \cite{tsu02}) we find an E(B--V) of 3.28. 
This column density is consistent with the total galactic column 
density of 1.7--1.9$\cdot$10$^{22}$ \sdens, calculated with the FTOOLS 
command ``nh'' based on data from Dickey \& Lockmann (\cite{dic90}). 
The equivalent logarithmic extinction is 4.94 and this value was used 
to correct all observed spectra for interstellar 
extinction. This, of course, assumes a constant extinction 
over the remnant but it is possible that the extinction 
may vary. We note here the presence of LDN 617 in the south areas 
of \gsnr\ (Lynds \cite{lyn62}). However, in the absence 
of detailed and accurate measurements we assume it is constant. 
It is evident from Table \ref{sfluxes} that the heavy extinction 
greatly affects the bluer lines making them very hard to detect.
\par
All spectra are characterized by the low but fairly constant \ha\ flux and 
the strong sulfur flux relative to \ha\ (\sii/\ha$\simeq$1.1--1.7).  
The ratio of the two sulfur lines approaches the low density limit in 
all cases and is estimated that the electron densities over the specific
locations are less than $\sim$200 \dens. 
Emission at \hbeta\ 4861 \AA\ is marginally present in one set of spectra
(location III) and may be a spurious detection given its low signal to noise.
The failure to detect the \hbeta\ line prevents us from checking on whether 
the spectra originate from fully developed or incomplete shock structures. 
In the following, we will assume that the radiative shock structures are
complete.
In addition, the medium ionization line of \oiii\ 5007 \AA\ appears to be absent
from the optical spectra. Emission at the 3$\sigma$ confidence level in this
line is detected only in one of the two spectra taken from position I 
suggesting a probable statistical fluctuation (Table \ref{sfluxes}). 
The 3$\sigma$ upper limit on the \oiii\ emission in all
spectra is $\sim$0.4 \flux, roughly a factor of 10 less than the typical \ha\
flux. The \oii 6300, 6364 \AA\ line fluxes from Area III should be treated with 
caution since there are indications for improper background subtraction.
\par
Note here that the signal to noise ratios given in Table~\ref{sfluxes} do not 
incorporate errors due to the calibration process which are $\sim$10\%.
%--------------------------------------------------------
  \begin {figure}
  \resizebox{\hsize}{!}{\includegraphics{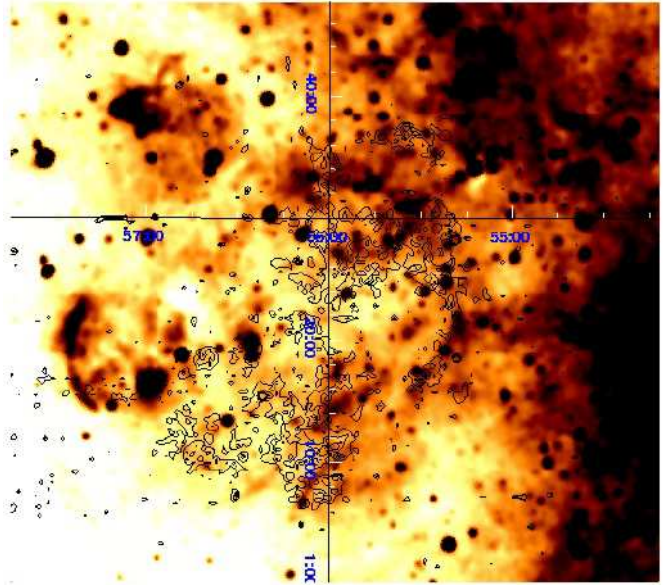}}
    \caption{Dust emission at 8.28 $\mu$m is shown as a greyscale image, while 
    the contours show the \hnii\ emission. The shadings run linearly 
    from  5.64 to 11.75 10$^{-14}$ erg s$^{-1}$ cm$^{-2}$ arcsec$^{-2}$ or
    equivalently from 2.4 to 5.0$\cdot$10$^{-6}$ W m$^{-2}$ sr$^{-1}$.
    Infrared emission detected in the western areas of the remnant 
    seems correlated with the optical emission} 
      \label{fig03}
% The optical contours scale linearly from 5 to 15 10^-17 erg/s/cm^2/sec/arcsec^2
% 1 W/m2/sr = 2.35E-8 erg/s/cm2/arcsec2
% 2.4 to 5.0$\cdot$10$^{-6}$ W m$^{-2}$ sr$^{-1}$
  \end{figure}
%
%--------------------------------------------------------
  \begin {figure*}
   \resizebox{\hsize}{!}{\includegraphics{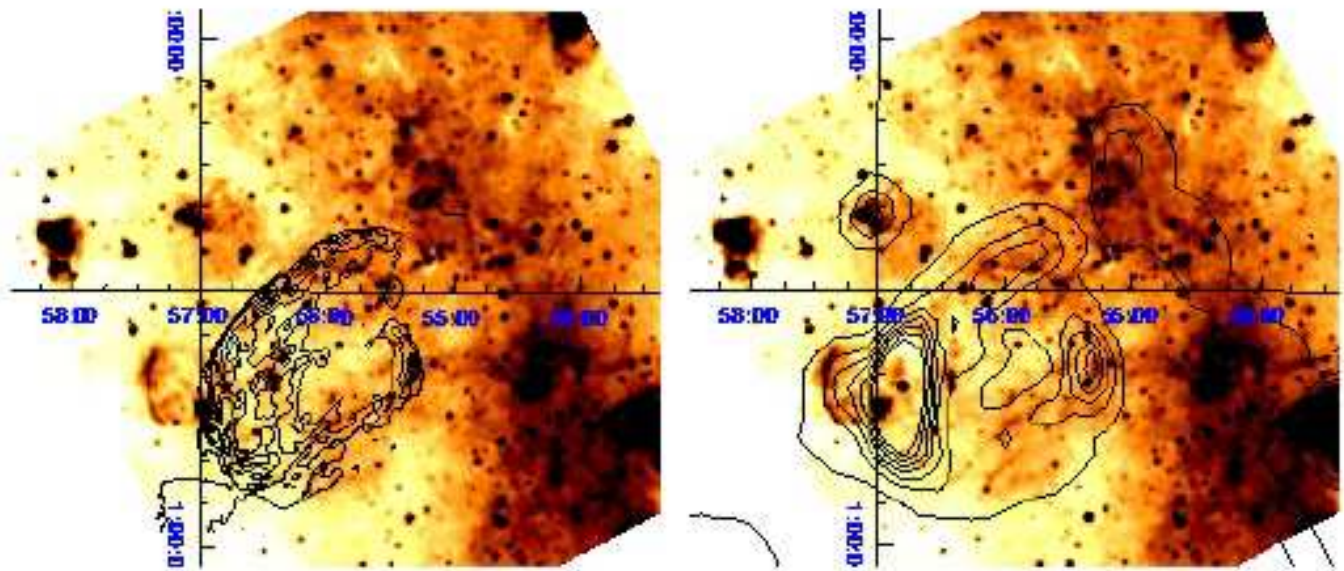}}
   \caption{ The MSX data at 8 $\mu$m are shown in this figure together with 
     contours of radio emission at 1442 MHz (15\arcsec\ resolution; left) 
     and 4850 MHz (3\arcmin.5 resolution; right).
     The 1442 MHz contours scale from 0.08 to 0.26 Jy/beam, while the 
     4850 MHz contours scale from 0.08 to 2.0 Jy/beam. 
     } 
     \label{fig04}
  \end{figure*}
%
% 1442 MHz 7 contours, 3 smoothing from 0.08 to 0.26
% 4850 MHz 7 contours, 3 smoothing from 0.08 to 2.0

%
\section{Discussion}
The known supernova remnant \gsnr\ lies in the galactic plane at a distance of
$\sim$2.5 kpc (Cox \et\ \cite{cox99}). As a result, short wavelengths suffer 
significant extinction rendering optical measurements a difficult task.
Earlier optical observations mainly focused on the morphology of the optical
emission and provided estimates of the fluxes in the \hnii\ and \sii\ lines
(Giacani \et\ \cite{gia97}). 
\par
In this work we present flux calibrated images of this faint remnant in 
major emission lines and long--slit spectra. 
The typical flux levels in the \hnii\ and \sii\ images are 
$\sim$8 and 4$\cdot$\flux, while emission at the bluer line of \oiii 5007\AA\ 
was not detected at the sensitivity limit of our observations. Note that 
for a color excess of 3.28 (see \S 4), the observed intensity of the \oiii\ 
line would be 30 times lower than that of the \ha\ line, even if they had 
equal intrinsic intensities. 
It is also evident from the typical fluxes of the low ionization lines that the
observed \sii\ flux\ is quite strong relative to \ha\ (see also \S 3.1).
The morphology of the optical
emission although diffuse and patchy (Figs. \ref{fig01} and \ref{fig02}) is very
well correlated with the radio emission, mainly in the west (Giacani \et\ 
\cite{gia97}). The nature of the optical emission may be attributed to
the presence of radiative shocks seen face--on due to the orientation 
of the interstellar ``clouds'' or to a patchy interstellar medium 
towards \gsnr. However, line--of--sight extinction variations cannot 
account for the overall appearance of the remnant, while it may not be 
clear why all these ``clouds'' should be oriented with this specific 
geometry. 
Another possibility 
may involve a turbulent magnetic field. The component parallel to the shock
front will be compressed not only in the transition region but also further 
in the recombination zone as the gas cools. The field component perpendicular 
to the shock front will not be affected and thus, there will be substantial  
variations of the magnetic pressure within the recombination zone. 
These variations will affect inversely the gas density distribution and 
consequently, the optical line emissivity will display a clumpy and 
irregular appearance (Raymond \& Curiel \cite{ray95}). 
\par
The optical spectra showing strong \sii, \nii\ and \ha\ emission and the 
calibrated optical images suggest that radiative shocks are present in W44. 
The neutral oxygen line at 6300\AA\ is detected although it is difficult 
to establish accurately the actual source flux because the night sky is 
very bright in this line. 
In an attempt to determine the physical conditions giving rise to the optical
emission we would have to explore the family of models appropriate for our
case (e.g. Cox \& Raymond \cite{cox85}, Raymond \et\ \cite{ray88}). Unfortunately, 
UV spectra and/or the \oiii 5007\AA\ and \oii 3727\AA\ lines are not available
and the actual family of radiative shocks cannot be confidently established.
The observed spectra are characterized by the strong \nii 6584 \AA, 
\sii 6716, 6731\AA\ line emission relative to \ha, the large sulfur line ratios 
implying low electron densities and the presence of moderate \oi 6300 \AA\ 
emission. The latter line is marginally present in 100 \vel\ shocks, while in
shocks faster than $\sim$150 \vel\ less flux in the \nii, \oi\ and \sii\ lines 
is produced. Thus, we estimate shock velocities in the range of 
110--150 \vel\ (Cox \& Raymond \cite{cox85}). The ionization state of the
preshocked gas cannot be resolved with the currrent data. The strong sulfur
emission and the low electron densities may suggest the presence of moderate
magnetic fields. Weak magnetic fields ($\sim$0.1--1 $\mu$G) allow for large
compression factors in the recombination zone, while magnetic fields 
higher than $\sim$5 $\mu$G suppress the compression resulting in lower 
electron and particle densities (e.g. Cox \cite{cox72}, Raymond \cite{ray79}, 
Raymond et al. \cite{ray88}).  
Adopting electron densities around 200 \dens\ and the estimated shock 
velocities, we find preshock cloud densities of the order of 3 atoms per 
cm$^3$ using the relation given by Fesen \& Kirshner (\cite{fes80}). 
However, this relation assumes negligible magnetic field and since it 
is probable that moderate magnetic fields are present what is obtained is 
actually a lower limit to the preshock could density. 
Shelton \et\ (\cite{she99}) estimate the velocity of the primary shock along 
the tenuous end to be $\sim$300 \vel. Assuming pressure equilibrium between the 
ISM and the interstellar clouds, a density contrast of $\sim$4--7 or greater 
is calculated. 
\par
Thermal emission from dust grains attributed to \gsnr\ has been detected in 
the infrared by the IRAS satellite (Arendt \cite{are89}, Giacani \et\ \cite{gia97}).
The dust grains can be collisionally heated  by the primary blast wave which 
then emit in the infrared. Emission was present in all four IRAS bands but the
moderate resolution of $\sim$2\arcmin\ did not allow a detailed comparison 
of the optical and/or radio data with the dust emission. 
However, the Midcourse Space Experiment (\footnote{The MSX data were 
retrieved from the NASA/IPAC Infrared science archive at 
http://irsa.ipac.caltech.edu}MSX) provided data of higher
sensitivity and spatial resolution ($\sim$20\arcsec) than the IRAS data. 
In Fig. \ref{fig03} we show the MSX image at 8.28 $\mu$m with the optical
emission in \hnii\ as the overlaid contours.  
Faint filamentary structures appear to correlate well with the optical 
emission. In fact, there is an arc--like structure in 
the infrared which is partially correlated with the optical and radio 
emission of
the remnant (Figs. \ref{fig03} and \ref{fig04}). The arc originates from 
\a\ $\simeq$18\h55\m33\s, and \dd\ $\simeq$ 01\degr15\arcmin20\arcsec, runs to
the north for $\sim$18\arcmin\  and then turns to the east up 
to \a\ $\simeq$18\h56\m13\s\ and \dd\ $\simeq$01\degr32\arcmin33\arcsec. The 
approximate radius of curvature of this arc is $\sim$11\arcmin. 
In addition, filamentary structures in the south also seem to match 
the radio emission at 1442 MHz (Fig. \ref{fig04}). 
These structures are a few arcminutes long and $\sim$0\arcmin.5 wide. 
If these structures are actually correlated to \gsnr\ then 
temperatures around 10$^6$ K would also be expected at the outer areas 
of the remnant.
Another very interesting feature is located further to the east at 
\a\ $\simeq$18\h57\m25\s\ and \dd\ $\simeq$01\degr18\arcmin15\arcsec\ extending
for $\sim$10\arcmin\ in the south--north direction. Even though, the high
resolution 1442 MHz data extend only up to 18\h57\m20\s, the lower
resolution 4850 MHz data match very well the position and curvature of this 
infrared filamentary structure. However, the nature of this structure 
(thermal or non--thermal) and its association to \gsnr\ cannot be 
reliably determined  using the currently available data. Dedicated radio
spectral observations are needed to determine whether the extent of the 
remnant is larger in the east--west direction than we currently 
assume or the observed emission is simply projected on \gsnr. 
The irregular boundary of the radio emission in the west, north--west
may be indicative of an interaction with the clouds in this area
(Fig. \ref{fig04}, and also Seta et al. \cite{set98}).
The interaction with molecular clouds in the east may be less extended
(e.g. Wootten \cite{woo77}) explaining the filamentary radio emission
in this area. However, the possibility of interaction in the west 
is an open issue (e.g. Cox \et\ \cite{cox99}) and more kinematic 
data would be needed.
For example, high resolution echelle spectra  around the remnant's
boundary would help to validate or discard this hypothesis  by
establishing the range of  velocities in the east and the west.
\section{Conclusions}
Weak and diffuse emission is the major characteristic of the field 
of the supernova remnant \gsnr. The long--slit spectra verify the 
shock heated nature of the detected radiation and suggest low 
electron densities. The velocities of the shocks travelling into the
interstellar clouds are in the range of 110--150 \vel. 
The morphology of the images and the low electron densities implied 
by the spectra may indicate the presence of moderate magnetic fields. 
High resolution infrared data are rather well correlated with the optical
and radio emission from \gsnr.  
\begin{acknowledgements}
The authors would like to thank the referee for his comments and suggestions 
and E. B. Giacani and J. J. Condon for 
providing the 1442 MHz and 4850 MHz digital data, respectively.
Skinakas Observatory is a collaborative project of the University of
Crete, the Foundation for Research and Technology-Hellas and
the Max-Planck-Institut f\"ur Extraterrestrische Physik. 
This research made use of data products from the Midcourse Space 
Experiment.  Processing of the data was funded by the Ballistic 
Missile Defense Organization with additional support from NASA 
Office of Space Science.  This research has also made use of the 
NASA/ IPAC Infrared Science Archive, which is operated by the 
Jet Propulsion Laboratory, California Institute of Technology, 
under contract with the National Aeronautics and Space 
Administration.
\end{acknowledgements}
%

%To match the observed spectra I found the following 
%AreaI   : Vs=110 km/s log(n)=0.5    B=7 microG
%AreaII  : Vs=140 km/s log(n)=0.7    B=5 microG
%AreaIII : Vs=100 km/s log(n)=0.9395 B=10 microG
%
\end{document}